\documentclass[a4paper]{article}

\usepackage{INTERSPEECH2018}
\usepackage{multirow}
\usepackage{CJK}
\usepackage[whole]{bxcjkjatype}
\usepackage[unicode]{hyperref}

\title{A Comparison of Modeling Units in Sequence-to-Sequence Speech Recognition with the Transformer on Mandarin Chinese}
\name{Shiyu Zhou$^1$$^,$$^2$, Linhao Dong$^1$$^,$$^2$, Shuang Xu$^1$, Bo Xu$^1$}
\address{
  $^1$Institute of Automation, Chinese Academy of Sciences \\
  $^2$University of Chinese Academy of Sciences }
\email{\{zhoushiyu2013, donglinhao2015, shuang.xu, xubo\}@ia.ac.cn}

\begin{document}
\begin{CJK*}{UTF8}{song}

\maketitle
\begin{abstract}
The choice of modeling units is critical to automatic speech recognition (ASR) tasks. Conventional ASR systems typically choose context-dependent states (CD-states) or context-dependent phonemes (CD-phonemes) as their modeling units.
However, it has been challenged by sequence-to-sequence attention-based models, which integrate an acoustic, pronunciation and language model into a single neural network. On English ASR tasks, previous attempts have already shown that the modeling unit of graphemes can outperform that of phonemes by sequence-to-sequence attention-based model.

In this paper, we are concerned with modeling units on Mandarin Chinese ASR tasks using sequence-to-sequence attention-based models with the Transformer. Five modeling units are explored including context-independent phonemes (CI-phonemes), syllables, words, sub-words and characters.
Experiments on HKUST datasets demonstrate that the lexicon free modeling units can outperform lexicon related modeling units in terms of character error rate (CER). Among five modeling units, character based model performs best and establishes a new state-of-the-art CER of \emph{$26.64\%$} on HKUST datasets without a hand-designed lexicon and an extra language model integration, which corresponds to a \emph{$4.8\%$} relative improvement over the existing best CER of $28.0\%$ by the joint CTC-attention based encoder-decoder network.

\end{abstract}

\section{Introduction}

Conventional ASR systems consist of three independent components: an acoustic model (AM), a pronunciation model (PM) and a language model (LM), all of which are trained independently. CD-states and CD-phonemes are dominant as their modeling units in such systems \cite{dahl2012context,sak2014long,senior2015context}. However, it recently has been challenged by sequence-to-sequence attention-based models.
These models are commonly comprised of an \emph{encoder}, which consists of multiple recurrent neural network (RNN) layers that model the acoustics, and a \emph{decoder}, which consists of one or more RNN layers that predict the output sub-word sequence. An \emph{attention} layer acts as the interface between the encoder and the decoder: it selects frames in the encoder representation that the decoder should attend to in order to predict the next sub-word unit \cite{prabhavalkar2017analysis}. In \cite{sainath2017no}, Tara et al. experimentally verified that the grapheme-based sequence-to-sequence attention-based model can outperform the corresponding phoneme-based model on English ASR tasks. This work is very interesting and amazing since a hand-designed lexicon might be removed from ASR systems. As we known, it is very laborious and time-consuming to generate a pronunciation lexicon. Without a hand-designed lexicon, the design of ASR systems would be simplified greatly. Furthermore, the latest work shows that attention-based encoder-decoder architecture achieves a new state-of-the-art WER on a $12500$ hour English voice search task using  the word piece models (WPM), which are sub-word units ranging from graphemes all the way up to entire words \cite{chiu2017state}.

Since the outstanding performance of grapheme-based modeling units on English ASR tasks, we conjecture that maybe there is no need for a hand-designed lexicon on Mandarin Chinese ASR tasks as well by sequence-to-sequence attention-based models.
In Mandarin Chinese, if a hand-designed lexicon is removed, the modeling units can be words, sub-words and characters.
Character-based sequence-to-sequence attention-based models have been investigated on Mandarin Chinese ASR tasks in \cite{chan2016online,shanattention}, but the performance comparison with different modeling units are not explored before. Building on our work \cite{2018arXiv180410752Z}, which shows that syllable based model with the Transformer can perform better than CI-phoneme based counterpart, we investigate five modeling units on Mandarin Chinese ASR tasks, including CI-phonemes, syllables (pinyins with tones), words, sub-words and characters. The Transformer is chosen to be the basic architecture of sequence-to-sequence attention-based model in this paper \cite{2018arXiv180410752Z,speechtransformer}.
Experiments on HKUST datasets confirm our hypothesis that the lexicon free modeling units, i.e. words, sub-words and characters, can outperform lexicon related modeling units, i.e. CI-phonemes and syllables. Among five modeling units, character based model with the Transformer achieves the best result and establishes a new state-of-the-art CER of \emph{$26.64\%$} on HKUST datasets without a hand-designed lexicon and an extra language model integration, which is a \emph{$4.8\%$} relative reduction in CER compared to the existing best CER of $28.0\%$ by the joint CTC-attention based encoder-decoder network with a separate RNN-LM integration \cite{hori2017advances}.

The rest of the paper is organized as follows. After an overview of the related work in Section \ref{label_related work}, Section \ref{label_system_overview} describes the proposed method in detail. we then show experimental results in Section \ref{label_experiment} and conclude this work in Section \ref{label_conclusions}.

\section{Related work}
\label{label_related work}

Sequence-to-sequence attention-based models have achieved promising results on English ASR tasks and various modeling units have been studied recently, such as CI-phonemes, CD-phonemes, graphemes and WPM \cite{prabhavalkar2017analysis,sainath2017no,chiu2017state,prabhavalkar2017comparison}.
In \cite{sainath2017no}, Tara et al. first explored sequence-to-sequence attention-based model trained with phonemes for ASR tasks and compared the modeling units of graphemes and phonemes. They experimentally verified that the grapheme-based sequence-to-sequence attention-based model can outperform the corresponding phoneme-based model on English ASR tasks.
Furthermore, the modeling units of WPM have been explored in \cite{chiu2017state}, which are sub-word units ranging from graphemes all the way up to entire words. It achieved a new state-of-the-art WER on a $12500$ hour English voice search task.

Although sequence-to-sequence attention-based models perform very well on English ASR tasks, related works are quite few on Mandarin Chinese ASR tasks. Chan et al. first proposed Character-Pinyin sequence-to-sequence attention-based model on Mandarin Chinese ASR tasks. The Pinyin information was used during training for improving the performance of the character model. Instead of using joint Character-Pinyin model, \cite{shanattention} directly used Chinese characters as network output by mapping the one-hot character representation to an embedding vector via a neural network layer. What's more, \cite{zou2018comparable} compared the modeling units of characters and syllables by sequence-to-sequence attention-based models.

Besides the modeling unit of character, the modeling units of words and sub-words are investigated on Mandarin Chinese ASR tasks in this paper. Sub-word units encoded by byte pair encoding (BPE) have been explored on neural machine translation (NMT) tasks to address out-of-vocabulary (OOV) problem on open-vocabulary translation \cite{sennrich2015neural}, which iteratively replace the most frequent pair of characters with a single, unused symbol. We extend it to Mandarin Chinese ASR tasks. BPE is capable of encoding an open vocabulary with a compact symbol vocabulary of variable-length sub-word units, which requires no shortlist.

\section{System overview}
\label{label_system_overview}

\subsection{ASR Transformer model architecture}
\label{label_transformer_model}

The Transformer model architecture is the same as sequence-to-sequence attention-based models except relying entirely on self-attention and position-wise, fully connected layers for both the encoder and decoder \cite{vaswani2017attention}. The encoder maps an input sequence of symbol representations \textbf{x} = $\left( x_1, ..., x_n \right)$ to a sequence of continuous representations \textbf{z} = $\left( z_1, ..., z_n \right)$. Given \textbf{z}, the decoder then generates an output sequence \textbf{y} = $\left( y_1, ..., y_m \right)$ of symbols one element at a time.

\begin{figure}[t]
        \centering
        \includegraphics[width=1.0\linewidth]{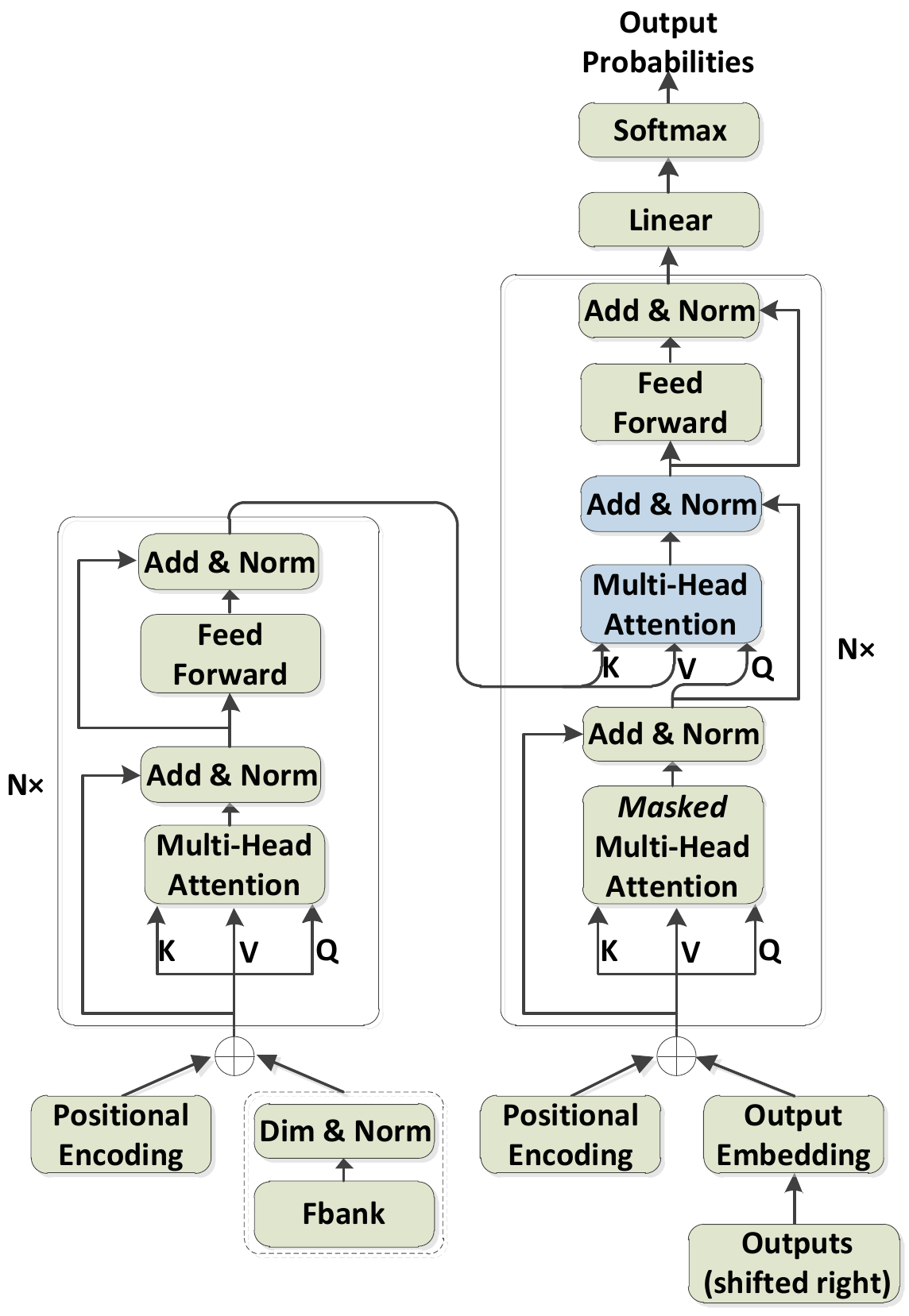}
        \caption{{\it The architecture of the ASR Transformer.}}
        \label{fig:fig_transformer}
\end{figure}

The ASR Transformer architecture used in this work is the same as our work \cite{2018arXiv180410752Z} which is shown in Figure~\ref{fig:fig_transformer}. It stacks multi-head attention (MHA) \cite{vaswani2017attention} and position-wise, fully connected layers for both the encode and decoder. The encoder is composed of a stack of $N$ identical layers. Each layer has two sub-layers. The first is a MHA, and the second is a position-wise fully connected feed-forward network. Residual connections are employed around each of the two sub-layers, followed by a layer normalization. The decoder is similar to the encoder except inserting a third sub-layer to perform a MHA over the output of the encoder stack. To prevent leftward information flow and preserve the auto-regressive property in the decoder, the self-attention sub-layers in the decoder mask out all values corresponding to illegal connections. In addition, positional encodings \cite{vaswani2017attention} are added to the input at the bottoms of these encoder and decoder stacks, which inject some information about the relative or absolute position of the tokens in the sequence.

The difference between the NMT Transformer \cite{vaswani2017attention} and the ASR Transformer is the input of the encoder. we add a linear transformation with a layer normalization to convert the log-Mel filterbank feature to the model dimension $d_{model}$ for dimension matching, which is marked out by a dotted line in Figure~\ref{fig:fig_transformer}.

\subsection{Modeling units}
\label{label_modeling_units}
Five modeling units are compared on Mandarin Chinese ASR tasks, including CI-phonemes, syllables, words, sub-words and characters.
Table~\ref{tab:different_modeling_units} summarizes the different number of output units investigated by this paper. We show an example of various modeling units in Table~\ref{tab:different_modeling_units_examples}.
        \begin{table}[th]
        \caption{\label{tab:different_modeling_units} {\it Different modeling units explored in this paper.}}
        \vspace{2mm}
        \centerline{
          \begin{tabular}{|c|c|}
            \hline
              {Modeling units}   & {Number of outputs}  \\
            \hline
            CI-phonemes &  $122$  \\
            \hline
            Syllables & $1388$  \\
            \hline
            Characters & $3900$  \\
            \hline
            Sub-words & $11039$  \\
            \hline
            Words & $28444$  \\
            \hline
          \end{tabular}
        }
        \end{table}

\subsubsection{CI-phoneme and syllable units}
\label{label_ci_syllable_units}

CI-phoneme and syllable units are compared in our work \cite{2018arXiv180410752Z}, which $118$ CI-phonemes without silence (phonemes with tones) are employed in the CI-phoneme based experiments and $1384$ syllables (pinyins with tones) in the syllable based experiments.
\emph{Extra tokens} (i.e. an unknown token (\textless UNK\textgreater), a padding token (\textless PAD\textgreater), and sentence start and end tokens (\textless S\textgreater/\textless \textbackslash S\textgreater)) are appended to the outputs, making the total number of outputs $122$ and $1388$ respectively in the CI-phoneme based model and syllable based model.
Standard tied-state cross-word triphone GMM-HMMs are first trained with maximum likelihood estimation to generate CI-phoneme alignments on training set. Then syllable alignments are generated through these CI-phoneme alignments according to the lexicon, which can handle multiple pronunciations of the same word in Mandarin Chinese.

The outputs are CI-phoneme sequences or syllable sequences during decoding stage. In order to convert CI-phoneme sequences or syllable sequences into word sequences, a greedy cascading decoder with the Transformer \cite{2018arXiv180410752Z} is proposed. First, the best CI-phoneme or syllable sequence $s$ is calculated by the ASR Transformer from observation $X$ with a beam size $\beta$. And then, the best word sequence $W$ is chosen by the NMT Transformer from the best CI-phoneme or syllable sequence $s$ with a beam size $\gamma$.
Through cascading these two Transformer models, we assume that $Pr(W|X)$ can be approximated.

Here the beam size $\beta=13$ and $\gamma=6$ are employed in this work.

\subsubsection{Sub-word units}
\label{label_subword_units}

Sub-word units, using in this paper, are generated by BPE \footnote{https://github.com/rsennrich/subword-nmt} \cite{sennrich2015neural}, which iteratively merges the most frequent pair of characters or character sequences with a single, unused symbol.
Firstly, the symbol vocabulary with the character vocabulary is initialized, and each word is represented as a sequence of characters plus a special end-of-word symbol `@@', which allows to restore the original tokenization. Then, all symbol pairs are counted iteratively and each occurrence of the most frequent pair (`A', `B') are replaced with a new symbol `AB'. Each merge operation produces a new symbol which represents a character n-gram. Frequent character n-grams (or whole words) are eventually merged into a single symbol. Then the final symbol vocabulary size is equal to the size of the initial vocabulary, plus the \emph{number of merge operations}, which is the hyperparameter of this algorithm \cite{sennrich2015neural}.

BPE is capable of encoding an open vocabulary with a compact symbol vocabulary of variable-length sub-word units, which requires no shortlist.
After encoded by BPE, the sub-word units are ranging from characters all the way up to entire words. Thus there are no OOV words with BPE and high frequent sub-words can be preserved.

In our experiments, we choose the \emph{number of merge operations} $5000$, which generates the number of sub-words units $11035$ from the training transcripts. After appended with $4$ \emph{extra tokens}, the total number of outputs is $11039$.

\subsubsection{Word and character units}
\label{label_word_char_units}

For word units, we collect all words from the training transcripts. Appended with $4$ \emph{extra tokens}, the total number of outputs is $28444$.

For character units, all Mandarin Chinese characters together with English words in training transcripts are collected, which are appended with $4$ \emph{extra tokens} to generate the total number of outputs $3900$ \footnote{we manually delete two tokens $\cdot$ and $+$, which are not Mandarin Chinese characters.}.

\section{Experiment}
\label{label_experiment}

\subsection{Data}
The HKUST corpus (LDC2005S15, LDC2005T32), a corpus of Mandarin Chinese conversational telephone speech, is collected and transcribed by Hong Kong University of Science and Technology (HKUST) \cite{liu2006hkust}, which contains 150-hour speech, and 873 calls in the training set and 24 calls in the test set. All experiments are conducted using 80-dimensional log-Mel filterbank features, computed with a 25ms window and shifted every 10ms. The features are normalized via mean subtraction and variance normalization on the speaker basis. Similar to \cite{sak2015fast,kannan2017analysis}, at the current frame $t$, these features are stacked with 3 frames to the left and downsampled to a 30ms frame rate. As in \cite{hori2017advances}, we generate more training data by linearly scaling the audio lengths by factors of $0.9$ and $1.1$ (speed perturb.), which can improve the performance in our experiments.

        \begin{table}[th]
        \caption{\label{tab:different_modeling_units_examples} {\it An example of various modeling units in this paper.}}
        \vspace{2mm}
        \centerline{
          \begin{tabular}{|c|c|}
            \hline
              {Modeling units}   & {Example}  \\
            \hline
            CI-phonemes &  Y IY1 JH UH3 NG3 X IY4 N4 N IY4 AE4 N4  \\
            \hline
            Syllables & YI1 ZHONG3 XIN4 NIAN4  \\
            \hline
            Characters & 一\ \ 种\ \ 信\ \ 念  \\
            \hline
            Sub-words & 一种\ \ 信@@\ \ 念  \\
            \hline
            Words & 一种\ \ 信念  \\
            \hline
          \end{tabular}
        }
        \end{table}

\subsection{Training}

We perform our experiments on the \emph{base model} and \emph{big model} (i.e. D512-H8 and D1024-H16 respectively) of the Transformer from \cite{vaswani2017attention}. The basic architecture of these two models is the same but different parameters setting. Table~\ref{tab:paramters} lists the experimental parameters between these two models. The Adam algorithm \cite{kingma2014adam} with gradient clipping and warmup is used for optimization. During training, label smoothing of value $\epsilon_{ls}=0.1$ is employed \cite{szegedy2016rethinking}. After trained, the last 20 checkpoints are averaged to make the performance more stable \cite{vaswani2017attention}.
        \begin{table}[th]
        \caption{\label{tab:paramters} {\it Experimental parameters configuration.}}
        \vspace{2mm}
        \centerline{
          \begin{tabular}{|c|c|c|c|c|c|c|}
            \hline
              {model}   & $N$  &    $d_{model}$ & $h$ & $d_k$ & $d_v$   & $warmup$  \\
            \hline
            D512-H8 &  $6$ &   $512$  & $8$ & $64$  &  $64$  & $4000\ steps$ \\
            \hline
            D1024-H16 & $6$ &   $1024$ & $16$ & $64$ &  $64$  & $12000\ steps$ \\
            \hline
          \end{tabular}
        }
        \end{table}
        
In the CI-phoneme and syllable based model, we cascade an ASR Transformer and a NMT Transformer to generate word sequences from observation $X$. However, we do not employ a NMT Transformer anymore in the word, sub-word and character based model, since the beam search results from the ASR Transformer are already the Chinese character level. The total parameters of different modeling units list in Table~\ref{tab:parameters_num}.

        \begin{table}
          \caption{\label{tab:parameters_num} {\it Total parameters of different modeling units.}}
          \newcommand{\tabincell}[2]{\begin{tabular}{@{}#1@{}}#2\end{tabular}}
          \centering
          \begin{tabular}{|c|c|c|c|}
            \hline
              {model}   &   \tabincell{c}{D512-H8\\ (ASR)}   &    \tabincell{c}{D1024-H16\\ (ASR)}  &  \tabincell{c}{D512-H8\\ (NMT)}  \\
            \hline
              CI-phonemes &  $57M$ &   $227M$  & $71M$  \\
            \hline
              Syllables &  $58M$ &   $228M$  & $72M$  \\
            \hline
              Words &  $71M$ &   $256M$  & $-$  \\
            \hline
              Sub-words &  $63M$ &   $238M$  & $-$  \\
            \hline
              Characters &  $59M$ &   $231M$  & $-$  \\
            \hline
          \end{tabular}
        \end{table}

\subsection{Results}

According to the description from Section \ref{label_modeling_units}, we can see that the modeling units of words, sub-words and characters are lexicon free, which do not need a hand-designed lexicon. On the contrary, the modeling units of CI-phonemes and syllables need a hand-designed lexicon.

Our results are summarized in Table~\ref{tab:modeling_units results}.
It is clear to see that the lexicon free modeling units, i.e. words, sub-words and characters, can outperform corresponding lexicon related modeling units, i.e. CI-phonemes and syllables on HKUST datasets. It confirms our hypothesis that we can remove the need for a hand-designed lexicon on Mandarin Chinese ASR tasks by sequence-to-sequence attention-based models.
What's more, we note here that the sub-word based model performs better than the word based counterpart. It represents that the modeling unit of sub-words is superior to that of words, since sub-word units encoded by BPE have fewer number of outputs and without OOV problems.
However, the sub-word based model performs worse than the character based model.
The possible reason is that the modeling unit of sub-words is bigger than that of characters which is difficult to train.
We will conduct our experiments on larger datasets and compare the performance between the modeling units of sub-words and characters in future work.
Finally, among five modeling units, character based model with the Transformer achieves the best result. It demonstrates that the modeling unit of character is suitable for Mandarin Chinese ASR tasks by sequence-to-sequence attention-based models, which can simplify the design of ASR systems greatly.

        \begin{table}[th]
        \caption{\label{tab:modeling_units results} {\it Comparison of different modeling units with the Transformer on HKUST datasets in CER (\%).}}
        \vspace{2mm}
        \centerline{
          \begin{tabular}{|c|c|c|}
            \hline
              {Modeling units}   & {Model}  &    {CER}   \\
            \hline
            \multirow{3} * {CI-phonemes \cite{2018arXiv180410752Z}} & D512-H8 &   $32.94$  \\
            & D1024-H16 &   \textbf{30.65}  \\
            & D1024-H16 (speed perturb) &   $30.72$  \\
            \cline{2-3}
            \hline
            \multirow{3} * {Syllables \cite{2018arXiv180410752Z}} & D512-H8 &   $31.80$  \\
            & D1024-H16 &   $29.87$  \\
            & D1024-H16 (speed perturb) &   \textbf{28.77}  \\
            \cline{2-3}
            \hline
            \multirow{3} * {Words} & D512-H8 &   $31.98$  \\
            & D1024-H16 &   $28.74$  \\
            & D1024-H16 (speed perturb) &   \textbf{27.42}  \\
            \cline{2-3}
            \hline
            \multirow{3} * {Sub-words} & D512-H8 &   $30.22$  \\
            & D1024-H16 &   $28.28$  \\
            & D1024-H16 (speed perturb) &   \textbf{27.26}  \\
            \cline{2-3}
            \hline
            \multirow{3} * {Characters} & D512-H8 &   $29.00$  \\
            & D1024-H16 &   $27.70$  \\
            & D1024-H16 (speed perturb) &   \textbf{26.64}  \\
            \cline{2-3}
            \hline
          \end{tabular}
        }
        \end{table}

\subsection{Comparison with previous works}

In Table~\ref{tab:comparison_with_previous}, we compare our experimental results to other model architectures from the literature on HKUST datasets.
First, we can find that our best results of different modeling units are comparable or superior to the best result by the deep multidimensional residual learning with 9 LSTM layers \cite{zhao2016multidimensional}, which is a hybrid LSTM-HMM system with the modeling unit of CD-states. We can observe that the best CER \emph{$26.64\%$} of character based model with the Transformer on HKUST datasets achieves a \emph{$13.4\%$} relative reduction compared to the best CER of $30.79\%$ by the deep multidimensional residual learning with 9 LSTM layers. It shows the superiority of the sequence-to-sequence attention-based model compared to the hybrid LSTM-HMM system.

Moreover, we can note that our best results with the modeling units of words, sub-words and characters are superior to the existing best CER of $28.0\%$ by the joint CTC-attention based encoder-decoder network with a separate RNN-LM integration \cite{hori2017advances}, which is the state-of-the-art on HKUST datasets to the best of our knowledge.
Character based model with the Transformer establishes a new state-of-the-art CER of \emph{$26.64\%$} on HKUST datasets without a hand-designed lexicon and an extra language model integration, which is a \emph{$7.8\%$} relative reduction in CER compared to the CER of \emph{$28.9\%$} of the joint CTC-attention based encoder-decoder network when no external language model is used, and a \emph{$4.8\%$} relative reduction in CER compared to the existing best CER of $28.0\%$ by the joint CTC-attention based encoder-decoder network with separate RNN-LM \cite{hori2017advances}.

      \begin{table}[th]
      \newcommand{\tabincell}[2]{\begin{tabular}{@{}#1@{}}#2\end{tabular}}
        \caption{\label{tab:comparison_with_previous} {\it CER (\%) on HKUST datasets compared to previous works.}}
        \vspace{2mm}
        \centerline{
          \begin{tabular}{|c|c|}
            \hline
              {model}   &    {CER}   \\
            \hline
            \tabincell{c}{LSTMP-9$\times$800P512-F444 \cite{zhao2016multidimensional}\ } &    $30.79$  \\
            \tabincell{c}{CTC-attention+joint dec. (speed perturb., one-pass) \\ +VGG net \\ +RNN-LM (separate) \cite{hori2017advances}\ } &    \tabincell{c}{ \\ $28.9$ \\ \textbf{28.0} }  \\
            \hline
             CI-phonemes-D1024-H16 \cite{2018arXiv180410752Z} &    $30.65$  \\
             Syllables-D1024-H16 (speed perturb) \cite{2018arXiv180410752Z} &    $28.77$  \\
             Words-D1024-H16 (speed perturb) &    $27.42$  \\
             Sub-words-D1024-H16 (speed perturb) &    $27.26$  \\
             Characters-D1024-H16 (speed perturb) &    \textbf{26.64}  \\
            \hline
          \end{tabular}
        }
      \end{table}

\section{Conclusions}
\label{label_conclusions}

In this paper we compared five modeling units on Mandarin Chinese ASR tasks by sequence-to-sequence attention-based model with the Transformer, including CI-phonemes, syllables, words, sub-words and characters. We experimentally verified that the lexicon free modeling units, i.e. words, sub-words and characters, can outperform lexicon related modeling units, i.e. CI-phonemes and syllables on HKUST datasets. It represents that maybe we can remove the need for a hand-designed lexicon on Mandarin Chinese ASR tasks by sequence-to-sequence attention-based models. Among five modeling units, character based model achieves the best result and establishes a new state-of-the-art CER of \emph{$26.64\%$} on HKUST datasets without a hand-designed lexicon and an extra language model integration, which corresponds to a \emph{$4.8\%$} relative improvement over the existing best CER of $28.0\%$ by the joint CTC-attention based encoder-decoder network. Moreover, we find that sub-word based model with the Transformer, encoded by BPE, achieves a promising result, although it is slightly worse than character based counterpart.



\section{Acknowledgements}

The authors would like to thank Chunqi Wang and Feng Wang for insightful discussions.

\bibliographystyle{IEEEtran}

\bibliography{mybib}

\end{CJK*}
\end{document}